# Zero-Field Thermal Hall Effect in Insulator

**Authors:**

Zhe Cui[1,2#], Haoran Fan[1,2#], Wenjiang Zhou[3,4#], Xianghong Jin[1,5], Yuchen Gu[1], Da Ma[6], Cong Xiao[7], Hua Jiang[7,8], Xincheng Xie[1,7,9], Bai Song[3,4*], Yuan Li[1,10*], and Xi Lin[1,9,11*]

**Affiliations:**

[1]International Center for Quantum Materials, School of Physics, Peking University; Beijing, 100871, China.

[2]Beijing Academy of Quantum Information Sciences; Beijing, 100193, China.

[3]National Key Laboratory of Advanced Micro and Nano Manufacture Technology, Peking University; Beijing, 100871, China.

[4]School of Mechanics and Engineering Science, Peking University; Beijing 100871, China.

[5]Liaoning Academy of Materials; Shenyang, 110167, China.

[6]College of Sciences, Northeastern University; Shenyang, 110819, China.

[7]Interdisciplinary Center for Theoretical Physics and Information Sciences, Fudan University; Shanghai, 200433, China.

[8]State Key Laboratory of Surface Physics, Institute for Nanoelectronic Devices and Quantum Computing, Fudan University; Shanghai, 200433, China.

[9]Hefei National Laboratory; Hefei, 230088, China.

[10]Beijing National Laboratory for Condensed Matter Physics, Institute of Physics, Chinese Academy of Sciences; Beijing, 100190, China

[11]Interdisciplinary Institute of Light-Element Quantum Materials and Research Center for Light-Element Advanced Materials, Peking University; Beijing, 100871, China.

[#]These authors contributed equally to this work.

[*]Corresponding author. Email: Bai Song (songbai@pku.edu.cn); Yuan Li (yuan.li@iphy.ac.cn); Xi Lin (xilin@pku.edu.cn).

**Abstract:**

Fourier's law dictates that heat flow is usually parallel to the applied temperature gradient. However, under a high magnetic field, heat flow carried by both electrons in conductors and phonons in insulators can be deflected, a phenomenon known as thermal Hall effect. Intriguingly, we observe at zero field a spontaneous thermal Hall effect in an antiferromagnetic insulator. Despite a vanishingly small uncompensated magnetization, the magnitude of this effect is surprisingly large, comparable to typical responses induced by several teslas of external field. This zero-field behavior indicates that charge-neutral heat carriers can be governed by an intrinsic effective field arising from the unique spin arrangement. Our discovery challenges the centuries-old preconception of heat conduction and open up new avenues for exploring non-trivial topological responses in quantum materials.

**Main:**

The concept of condensed matter physics has been repeatedly expanded by diverse unconventional Hall responses. In particular, the unexpected experimental discoveries of the integer and fractional quantum Hall effects[1,2] unveiled an entirely new world of topological orders and fractionalized excitations. These intriguing phenomena in electronic systems not only serve as a definitive probe of topological states[3,4], but also provide a rigorous platform for exploring many-body electrons[5,6]. Recently, the emerging observations of thermal Hall effect (THE) have further enriched this realm of fundamental science[7–11]. Whereas THE mediated by charge carriers can often be attributed to the Wiedemann-Franz law[12], the microscopic origins of THE in insulators are substantially more complex and remain intensely debated. For example, phonons may give rise to Hall-like behaviors through several mechanisms including non-trivial Berry curvature, asymmetric extrinsic scattering, and coupling with other collective excitations[13–17]. By probing the transverse temperature gradient arising from charge-neutral excitations, THE in insulators opens a unique window into diverse quantum degrees of freedom not accessible by electrical transport, such as non-Abelian charge-neutral modes[18] and fractionalization of quantum spins[19].

Thermal Hall effect, similar to its electronic counterpart, is deeply rooted in the breaking of time-reversal symmetry (TRS) and therefore usually requires an external magnetic field as a prerequisite. Notably, in all existing reports of THE in insulators[13–17,20–24], the Hall angle remains minute at ~1‰ even under strong fields of ~10 teslas, highlighting the insensitivity of heat-carrying neutral excitations to TRS breaking. This pronounced insensitivity naturally poses a great conceptual challenge when considering a spontaneous thermal Hall effect (STHE) without an external field, in stark contrast to carrier transport where charge can be readily redirected by the intrinsic magnetic landscape of a material[25–27]. Indeed, to experimentally observe STHE, an exceptionally strong effective magnetic field would be indispensable, in conjunction with extremely reliable thermal measurements.

Here, we explore the thermal Hall effect in $Na_2Co_2TeO_6$—an emerging antiferromagnetic insulator. The material is first cooled into its magnetically ordered phase under a training field. Subsequently, the field is removed and a longitudinal heat flow is applied under zero field to initiate thermal transport measurements. Remarkably, we observe the emergence of a large spontaneous temperature gradient in the transverse direction. The magnitude of this transverse response is comparable to typical thermal Hall signals measured under multi-tesla external fields, and precludes a straightforward explanation based on uncompensated magnetization, which is not only vanishingly small but also exhibits a distinct temperature dependence. Instead, our analysis of the underlying spin texture suggests an effective intrinsic field that is surprisingly strong and can therefore deflect the charge-neutral heat carriers.

**Experimental Setup**

To experimentally probe thermal Hall effect in the layered insulator $Na_2Co_2TeO_6$, a millimeter-sized and micrometer-thick (90 μm) single-crystal sample is mounted in a cryostat as illustrated in Fig. 1a, together with the corresponding crystallographic axes (Supplementary Fig. 1). Below its Néel temperature $T_N \approx 27$ K[28], $Na_2Co_2TeO_6$ exhibits non-coplanar antiferromagnetism with a distinctive spin texture comprising in-plane vortices $v$ and nearly compensated out-of-plane moments $m$[29,30]. We prepare a single antiferromagnetic domain (concerning magnetic point group) by cooling the sample from above $T_N$ to a base temperature $T_{base}$ under a weak magnetic field $B$ along the $c$-axis and then removing the field (Fig. 1b). Notably, the state prepared under a negative field ($-B_{train}$) is simply the time-reversal conjugate of that under a positive field of equal magnitude ($+B_{train}$)[30].

With the sample properly trained, a heat flow $\dot{Q}$ is then generated along the $a$-axis via an electrical heater, which always induces a longitudinal temperature difference $\Delta T_{xx}$ according to Fourier's law. If THE exists, a transverse temperature difference $\Delta T_{xy}$ along the $a^*$-axis will also appear. Photographs of the measurement setup are provided in Supplementary Fig. 2. Both $\Delta T_{xx}$ and $\Delta T_{xy}$ are recorded directly using a high-resolution differential thermocouple configuration (Supplementary Figs. 3 and 4), wherein heat leakage (Supplementary Fig. 5) and signal mixing (Supplementary Fig. 6) at the contacts are negligible. The sign convention for $\Delta T_{xy}$ follows the right-hand rule with respect to the heat flow $\dot{Q}$ and the magnetic field $B$, consistent with the convention for the electrical Hall effect.

In Fig. 1c, we present the raw Seebeck voltages and the corresponding temperature differences measured in field-trained $Na_2Co_2TeO_6$ at 23 K. As expected, we obtain a $\Delta T_{xx}$ that remains symmetric under time reversal. In addition, we also observe a finite $\Delta T_{xy}$, which may have arisen trivially from the cross-talk between the longitudinal and transverse directions. However, a closer comparison of $\Delta T_{xy}^{+}$ and $\Delta T_{xy}^{-}$ (corresponding to $+B_{train}$ and $-B_{train}$, respectively) reveals a remarkable asymmetry upon time reversal well above our resolution limit of ~6 nV (~0.6 mK), which is an unequivocal manifestation of THE. In comparison, no discernible signal was observed above $T_N$ (for example at 30 K) within the experimental uncertainty (Supplementary Fig. 7). By employing antisymmetrization, the genuine Hall response can be readily extracted as $(\Delta T_{xy}^{+} - \Delta T_{xy}^{-})/2$. This operation cancels any artifacts induced by sample geometric asymmetry and thermocouple misalignment inherent to a handmade device.

**Thermal Hall Effect at Zero Magnetic Field**

We then proceed to extract the zero-field thermal conductivities (Eqs. 1–4) of $Na_2Co_2TeO_6$ over a wide temperature range from 2 to 90 K. For the longitudinal thermal conductivity, $\kappa_{xx}$, we observe a temperature dependence that is characteristic of phonon-mediated transport in high quality crystalline insulators, with a pronounced peak of approximately 12 W $K^{-1}$ $m^{-1}$ at $T_p \approx 47$ K (Fig. 2b, inset). In contrast, the transverse thermal conductivity, $\kappa_{xy}$ (Fig. 2b), is only on the order of 1 mW $K^{-1}$ $m^{-1}$. Despite its much smaller magnitude, $\kappa_{xy}$ is unlikely to arise from spurious leakage of the longitudinal signal. This is first guaranteed by antisymmetrization over two spin configurations of the same sample that are related by time-reversal symmetry, and is further supported by the distinct temperature dependence of $\kappa_{xy}$ compared to $\kappa_{xx}$. Indeed, a key feature of the transverse thermal response is its correspondence with the magnetic characteristics (Fig. 2a). Since all measurements are performed under zero external magnetic field, we therefore reveal that TRS breaking in the spin arrangements alone is sufficient to generate a STHE.

A spontaneous thermal Hall signal becomes observable below the Néel temperature $T_N$ (Fig. 2b). Intriguingly, $\kappa_{xy}$ exhibits a complex trend with temperature rising characterized by negative values, positive values, and a change of sign, indicating a strong correlation with the temperature-driven evolution of spin configurations (Fig. 2a). To maintain consistency with the sign convention commonly adopted for the thermal Hall effect in insulating systems under magnetic fields, we plot $-\kappa_{xy}$ instead of $\kappa_{xy}$. As temperature rises from 2 K, $-\kappa_{xy}$ evolves from a positive value to a negative one, undergoing a sign reversal at the compensation point $T^*$ which is associated with the weak out-of-plane ferrimagnetism[31]. The magnitude of $-\kappa_{xy}$ exhibits a positive maximum of approximately 0.7 mW $K^{-1}$ $m^{-1}$ at 6.5 K and a negative extremum of around 2.2 mW $K^{-1}$ $m^{-1}$ at 23 K. As the temperature exceeds 23 K and further increases toward $T_N$, the thermal Hall signal rapidly diminishes and approaches zero. The absence of a detectable thermal Hall response above $T_N$ further suggests that the spontaneous signal observed at lower temperatures originates from spin configurations, rather than from extrinsic factors.

Control experiments performed under different training fields and heating powers reproduce the key features of the STHE described above (Supplementary Figs. 8 and 9). Moreover, the zero-field thermal Hall effect is independently observed in a separate sample when the heat current $\dot{Q}$ is applied along the $a^*$-axis (Supplementary Fig. 10). These measurements collectively demonstrate the robustness of the observed effect.

**Spontaneous Thermal Hall Effect from a Tesla-Scale Effective Field**

The thermal Hall angle is defined as $\kappa_{xy}/\kappa_{xx}$. In the antiferromagnetic insulator $Na_2Co_2TeO_6$, its measured zero-field value reaches ~1‰, comparable to the thermal Hall angles reported in other candidate antiferromagnetic insulators, such as cuprate superconductors (3.6‰ at 15 T)[10], α-$RuCl_3$ (1‰ at 15 T)[21], and $Cu_3TeO_6$ (3‰ at 15 T)[22]. In stark contrast to our zero-field observation, all of these systems require a strong external magnetic field to break TRS. This difference suggests that the role of the uncompensated ferrimagnetic moment $m$ should be carefully examined.

According to existing experimental reports and theoretical models (Supplementary Table. 1; refs. 32, 33), a sizable thermal Hall angle in ferromagnetic systems typically requires a magnetic moment on the order of ~1 $\mu_B$ per formula unit. In contrast, the magnitude of $m$ in our system (~0.01 $\mu_B$ per formula unit) is far too small to account for the observed signal. Furthermore, following the convention of previous studies (refs. 9, 10, 18–21; Supplementary Figs. 11 and 12), we compare $\kappa_{xy}/T$ with the temperature dependence of $m$. A clear discrepancy is observed, exceeding the experimental uncertainty (Fig. 3a, inset). These results indicate that the STHE in $Na_2Co_2TeO_6$ cannot be explained by established TRS-breaking mechanisms, such as a strong external field or a large net magnetic moment.

This finding points to an intriguing alternative mechanism: The spin texture itself may induce an effective breaking of TRS for heat transport. Microscopically, the spin configuration in $Na_2Co_2TeO_6$ features not only an out-of-plane $m$ but also an in-plane vortex $v$, both of which reverse sign under time-reversal operation. The latter has been estimated to amount to a net-moment density on the order of ~0.1 $\mu_B$ per formula unit by magneto-optical measurements and, unlike $m$, it does not change sign over the temperature range below $T_N$[30]. We therefore consider the simplest multiplicative combination, $m \cdot v^2$, which exhibits the expected time-reversal-odd symmetry (Supplementary Fig. 13) and the sign-reversal behavior at $T^*$. Remarkably, the sign of $\kappa_{xy}/T$ reverses precisely when $m \cdot v^2$ changes sign (Fig. 3a, inset), and its magnitude scales linearly with $m \cdot v^2$ (Fig. 3a). These observations suggest that $m$ and $v$ act cooperatively to generate an effective magnetic field, $B_{\mathrm{eff}} \propto m \cdot v^2$, which gives rise to a Hall-like response in $\kappa_{xy}/T$. We emphasize that, although the observed STHE correlates well with $m \cdot v^2$ rather than $m$, the underlying physical mechanism remains to be fully established, which requires further insight into the microscopic spin texture of $Na_2Co_2TeO_6$. Nonetheless, our results already demonstrate a fundamental effect: A non-coplanar spin texture can deflect heat flow.

To better appreciate the considerable magnitude of the STHE, we compare it with conventional THE under an external magnetic field $B_{ext}$ at temperatures above $T_N$ (purple shade in Fig. 3b, inset), where long-range spin order is absent and $B_{ext}$ is the sole source of TRS breaking. In this regime, $\kappa_{xy}$ exhibits typical phonon-dominated behaviors, closely tracking $\kappa_{xx}$, with both peaking at $T_p \approx 47$ K. The field dependence of $-\kappa_{xy}$ shows that $\kappa_{xy}/T$ increases linearly with $B_{ext}$ at temperatures above $T_N$ (Fig. 3b), consistent with phonon-mediated THE observed in other systems (Supplementary Fig. 14). The slope $d(\kappa_{xy}/T)/dB_{ext}$ increases upon cooling and reaches a maximum near $T_N$ with a value of approximately 0.03 mW $K^{-1}$ $m^{-2}$ $T^{-1}$.

Strikingly, the dependence of $\kappa_{xy}/T$ on $B_{ext}$ above $T_N$ (Fig. 3b) closely resembles its dependence on $B_{eff}$ below $T_N$ (Fig. 3a). From this correspondence, we estimate $B_{eff}$ to be at least 3.5 T at 21 K. Notably, the STHE and conventional THE exhibit qualitatively different responses: When a positive external field acts as a linear-response magnetization field above $T_N$, it makes $\kappa_{xy}$ negative; however, when it acts as a training field which produces positive $m$ (hence positive $B_{eff}$) immediately below $T_N$, it leads to positive $\kappa_{xy}$. Again, both the large effective field and the unconventional sign of $\kappa_{xy}$ demonstrate that the spin-texture-induced STHE cannot be trivially attributed to net moments in the system.

Further, we investigate THE under $B_{ext}$ at temperatures below $T_N$, where STHE and conventional THE coexist (blue shade in Fig. 3b, inset). The observation of near-zero $\kappa_{xy}$ upon sign reversal indicates that the STHE signals induced by $B_{eff}$ can be canceled by the opposing responses induced by $B_{ext}$. This cancellation again suggests that the effective field arising from the spin texture is on the order of several teslas. Meanwhile, the external field modifies the spin texture, leading to complex behaviors in $\kappa_{xy}$ (Supplementary Fig. 15). Taken together, these two independent estimations lead to a central conclusion: A non-coplanar spin texture in an insulator can induce a remarkably strong breaking of TRS, to which charge-neutral heat flow is highly sensitive, producing a response equivalent to that induced by an external magnetic field on the order of several teslas.

**Discussion**

The nature of the charge-neutral heat carriers and the origin of TRS breaking in the zero-field thermal Hall response of $Na_2Co_2TeO_6$ deserve to be discussed in detail. Several observations suggest that phonons may be responsible. First, under an external magnetic field, the conventional THE in $Na_2Co_2TeO_6$ exhibits a peak in $\kappa_{xy}$ which coincides with that in $\kappa_{xx}$ (Fig. 3b, inset; Supplementary Fig. 16), a signature of phonon-dominated Hall transport[11]. Therefore, it is reasonable to assume that phonons could continue to dominate as the field is reduced to zero. Second, the dependence of the STHE on the effective field (Fig. 3a) closely resembles the field dependence of the conventional THE in $Na_2Co_2TeO_6$ (Fig. 3b) and also other nonmagnetic insulators[23,24], where the thermal Hall response is firmly validated within a phonon-mediated framework[34].

A possible microscopic picture of phonon-mediated STHE is then considered. Over the past two decades, a variety of mechanisms have been proposed to explain phonon THE driven by an external field, including spin-phonon interactions[35,36], phonon Berry curvature[37,38], skew scattering by rare earth impurities[15], and effects associated with ionic bonding or anisotropic charge distributions[24]. To understand STHE, we note that lattices hosting spin textures can inherently break TRS for phonons, allowing them to acquire a nonzero Berry curvature and thereby give rise to an intrinsic Hall response. In addition, hybridization between phonons and magnons may also contribute to a nonzero thermal Hall signal[17].

Further, extrinsic mechanisms, such as skew scattering of phonons by the scalar spin chirality (SSC) defined as $\mathbf{S}_i \cdot (\mathbf{S}_j \times \mathbf{S}_k)$, could play a pivotal role in the observed STHE, where $\mathbf{S}_i$ is the spin of the *i*th lattice site[39]. In this scenario, SSC arising from non-coplanar spin textures deforms the electronic many-body wave function in Mott insulators[40], which in turn induces an effective magnetic field. For $Na_2Co_2TeO_6$, our analysis suggests the form of $B_{\mathrm{eff}} \propto m \cdot v^2$ (Supplementary Fig. 17). The SSC can drive asymmetric scattering of chiral phonons with opposite angular momentum, ultimately leading to a pronounced STHE.

Meanwhile, the magnetic excitations in antiferromagnetic insulators should also be considered[8]. Magnetic moments at two inequivalent lattice sites (Supplementary Fig. 17) may generate magnetic excitations with opposite deflections at zero field. At the compensation point $T^*$, since the magnetic moments cancel each other out ($m$ = 0), their contributions may give rise to a vanishing STHE. Furthermore, it has been reported that THE may involve simultaneous contributions from phonons and magnetic excitations, with the former dominating at high temperatures and the latter prevailing at low temperatures[32]. Although several intrinsic and extrinsic mechanisms appear promising, we note that the exact microscopic process through which spin textures interact with heat carriers remains an open question.

Finally, our observations may pose a challenge to the current phenomenological description of THE. Previous theoretical prediction states that, as temperature approaches zero, $\kappa_{xy}/T$ tends towards a constant for gapless fermions, but approaches zero for gapped fermions and bosons[34]. However, our experimental results show that $\kappa_{xy}/T$ no longer exhibits an explicit temperature dependence but is instead dictated solely by the spin arrangement (Fig. 3a). Specifically, for conventional phonon THE (Fig. 3b), we have $\kappa_{xy}/T = \alpha(T) \cdot B$, whereas for STHE over the broad temperature range from 27 K down to 2 K (Fig. 3a), we reveal $\kappa_{xy}/T = \alpha_0 \cdot B_{\mathrm{eff}}(T)$. Here, $\alpha(T)$ varies with temperature while $\alpha_0$ is a constant. This discrepancy highlights the need for measurements at even lower temperatures and potentially challenges existing theories.

**Conclusion**

In summary, our experimental observation of a transverse heat flow in an insulator under zero magnetic field firmly establishes a new form of Hall effect, adding to a rich family of phenomena with a long track-record of breaking boundaries and stimulating new physics. Our analysis reveals the indispensable role of spin textures for deflecting charge-neutral heat carriers. This trainable, spontaneous, and robust thermal Hall behavior offers a new platform for exploring complex magnetic structures with non-zero topology. Moreover, in light of the atypical heat flow direction and the absence of magnetic field, the STHE holds unique promise for addressing heat dissipation challenges in spintronic circuits and quantum computing.

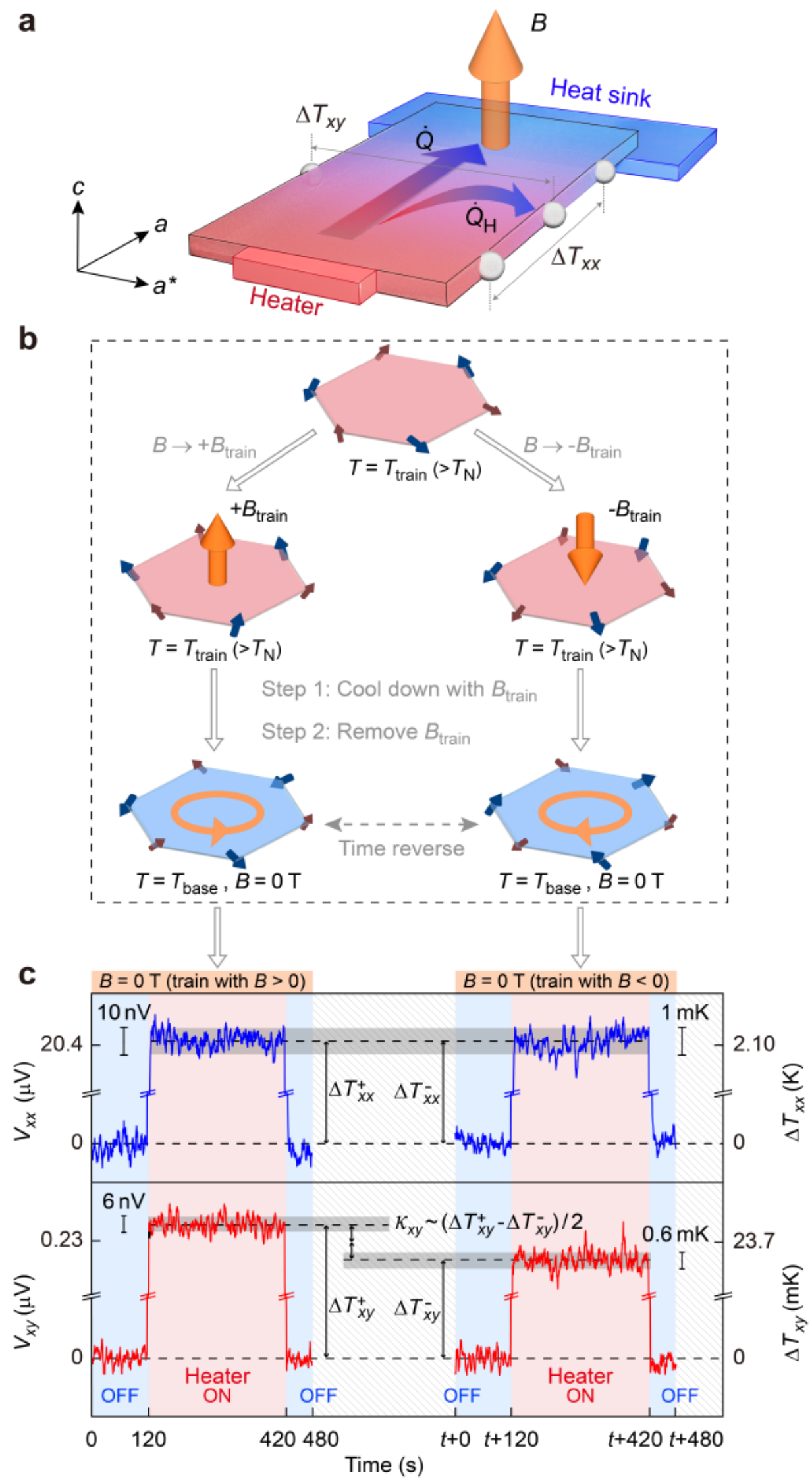

**Fig. 1 | Measurement of thermal Hall effect in $Na_2Co_2TeO_6$. a,** Schematic of the experimental setup. Heat flow is applied along the *a*-axis via a heater. The longitudinal and transverse temperature differences ($\Delta T_{xx}$ and $\Delta T_{xy}$) are measured by two sets of thermocouples, with the same sign convention of electrical Hall effect. An external magnetic field can be applied along the *c*-axis. **b,** Sample preparation procedure. First, a weak trained field $+B_{train}$ is applied with the sample temperature above its Néel temperature $T_N \approx 27$ K. The sample is then cooled down to $T_{base}$. Finally, the field is removed, leaving $Na_2Co_2TeO_6$ in an antiferromagnetic ground state with single domain (concerning magnetic point group). Training with $B < 0$ leads to the time-reversed counterpart of the ground state. **c,** Raw data of the STHE at 23 K. The top and bottom show $V_{xx}$ ($\Delta T_{xx}$) and $V_{xy}$ ($\Delta T_{xy}$), respectively, with noise levels around 10 nV (1 mK) and 6 nV (0.6 mK) shown by the gray color areas. The heating power is 1.5 mW.

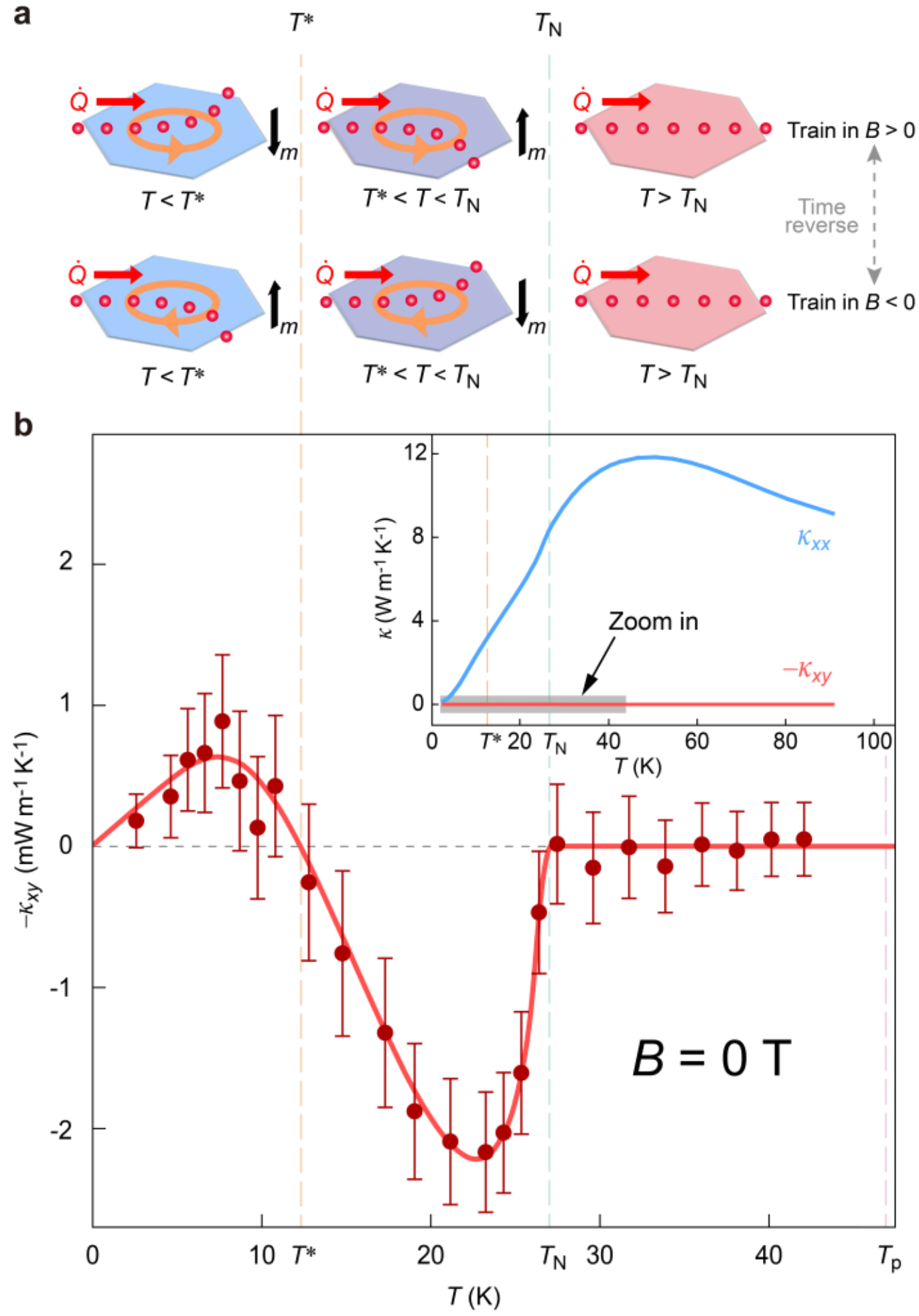

**Fig. 2 | Thermal Hall effect at zero magnetic field. a,** Schematic illustration of the spin arrangement and heat flow. For a given training process, the nearly compensated out-of-plane magnetization *m* and the in-plane vortices *v* act together to the deflection of heat flow. **b,** Temperature dependence of -$\kappa_{xy}$ at zero magnetic field. The solid line is a guide to the eyes. The error bars are obtained from the measurement noises, as shown the grey bar in Fig. 1c. Inset shows both $\kappa_{xx}$ and -$\kappa_{xy}$ in units of W m$^{-1}$ K$^{-1}$. Remarkably, reversal point of -$\kappa_{xy}$ coincides with the compensation point $T^*$ of *m*, while above $T_N$ there is no detectable -$\kappa_{xy}$, including the temperature $T_p$ of maximum $\kappa_{xx}$.

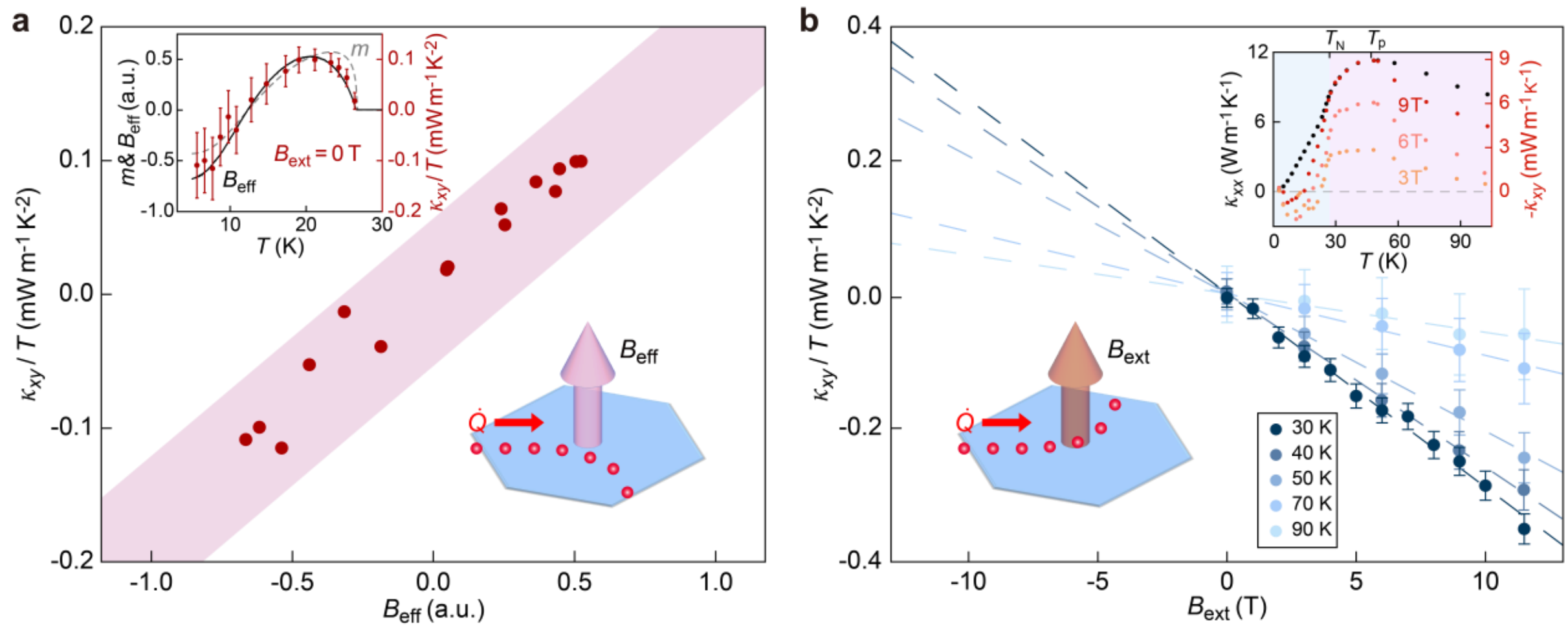

**Fig. 3 | Thermal Hall effect under effective magnetic field $B_{eff}$ and external magnetic field $B_{ext}$. a,** $\kappa_{xy}/T$ as a function of $B_{eff}$ at $B_{ext} = 0$ T, where $B_{eff} \propto m \cdot v^2$. Top-left inset: Comparison of $B_{eff}$ and $m$ with $\kappa_{xy}/T$. Bottom-right inset: Illustration of heat flow deflected by $B_{eff}$. **b,** $\kappa_{xy}/T$ as a function of $B_{ext}$ at different temperatures above $T_N$. The dashed lines are linear fittings. Top-right inset: $\kappa_{xx}$ under $B_{ext}$ = 9 T, -$\kappa_{xy}$ under $B_{ext}$ = 9 T, 6 T, and 3 T. For -$\kappa_{xy}$, two distinct regions are highlighted with blue and purple shades, which are separated by $T = T_N$. At $T < T_N$, THEs induced by $B_{eff}$ and $B_{ext}$ coexist; while at $T > T_N$, only $B_{ext}$ acts on phonons. Bottom-left inset: Illustration of phonon heat flow deflected by $B_{ext}$.

**Acknowledgments:**

We thank Junren Shi, Yang Liu, Jie Ma, Hongming Weng, Zhengmao Lu, Zhibin Gao, Jiansheng Wang, and Taekoo Oh for discussions. W.Z. acknowledges support from China Association for Science and Technology. B.S. acknowledges support from the New Cornerstone Science Foundation through the XPLORER PRIZE. We appreciate the High-performance Computing Platform of Peking University.

**Funding:**

National Key Research and Development Program of China Grant No. 2021YFA1401900 (X.L. and Y.L.), Grant No. 2025YFA1411500 (Y.L.)

Science Fund for Creative Research Groups from the National Natural Science Foundation of China Grant No. 52521007 (B.S.)

National Natural Science Foundation of China Grant No. 12141001 (X.L.), Grant No. 12474138 (Y.L.), Grant No. 525B2087 (W.Z.)

Quantum Science and Technology—National Science and Technology Major Project Grant No. 2021ZD0302600 (X.L.)

Scientific Research Innovation Capability Support Project for Young Faculty (ZYGXQNJSKYCXNLZCXM-E1) from the Ministry of Education of China (B.S.)

**Author contributions:**

X.L. and H.F. initiated the zero-field thermal Hall measurements. H.F. performed the measurement setup and discovered the phenomenon. Z.C. conducted the replicate experiments and verified the STHE phenomenon. Z.C., W.Z., and X.J. proposed that STHE originates from the spin texture. Z.C. prepared the manuscript with input from all authors. W.Z. performed the initial revision of the manuscript. X.J. provided the sample and its magnetic properties. H.F., X.J., Y.G., D.M., C.X., H.J., and X.X. reviewed the manuscript and provided constructive feedback. B.S., Y.L., and X.L. supervised this work.